\documentclass[11pt]{article}

\newsavebox{\junk}
\savebox{\junk}[1.6mm]{\hbox{$|\!|\!|$}}

\def\bbbz{{\mathchoice {\hbox{$\sf\textstyle Z\kern-0.4em Z$}}
{\hbox{$\sf\textstyle Z\kern-0.4em Z$}}
{\hbox{$\sf\scriptstyle Z\kern-0.3em Z$}}
{\hbox{$\sf\scriptscriptstyle Z\kern-0.2em Z$}}}}
\def\sq{\hbox{\rlap{$\sqcap$}$\sqcup$}}

\def\bbbr{{\rm I\!R}}

\usepackage{amsmath} 
\usepackage{amsfonts}
\usepackage{latexsym}
\usepackage{bm}

\newcommand{\field}[1]{\mathbb{#1}}

\newcommand{\unif}{\mbox{unif}}

\newcommand{\ben}{\begin{enumerate}}
\newcommand{\een}{\end{enumerate}}
\newcommand{\bit}{\begin{itemize}}
\newcommand{\eit}{\end{itemize}}

\newtheorem{theorem}{Theorem}[section]
\newtheorem{proposition}[theorem]{Proposition}

\newcommand{\ba}{\begin{array}{rcl}}
\newcommand{\ea}{\end{array}}
\newcommand{\bt}{\begin{theorem}}
\newcommand{\et}{\end{theorem}}
\newcommand{\bd}{\begin{description}}
\newcommand{\ed}{\end{description}}

\def\Expect{{\sf E}}

\def\slabel#1{\label{s:#1}}

\def\elabel#1{\label{e:#1}}
\def\eq#1/{(\ref{e:#1})}

\def\Section#1{Section~\ref{s:#1}}
\def\Re{\bbbr}

\def\beq{\begin{equation}}
\def\eeq{\end{equation}}
\def\beqa{\begin{eqnarray}}
\def\eeqa{\end{eqnarray}}

\def\qed{\ifmmode\sq\else{\unskip\nobreak\hfil
\penalty50\hskip1em\null\nobreak\hfil\sq
\parfillskip=0pt\finalhyphendemerits=0\endgraf}\fi}

\def\sqr#1#2{{\vcenter{\hrule height.#2pt
      \hbox{\vrule width.#2pt height#1pt \kern#1pt
         \vrule width.#2pt}
       \hrule height.#2pt}}}

\newcommand{\bc}{\begin{corollary}}
\newcommand{\ec}{\end{corollary}}

\newcommand{\bp}{\begin{proposition}}
\newcommand{\ep}{\end{proposition}}

\def\eye(#1){{\bf (#1)}\quad}

\def\taboo#1{{{}_{#1}}}

\def\0P{\taboo{0}P}
\def\0Pn{\taboo{0}P^n}

\usepackage{bbm}
\usepackage{amsmath}
\DeclareSymbolFont{largesymbols}{OMX}{yhex}{m}{n}
\DeclareMathAccent{\widewidehat}{\mathord}{largesymbols}{"62}

\usepackage{color}
\usepackage{graphicx}
\usepackage{subcaption}

\title{Perfect and $\varepsilon$-Perfect Simulation Methods for the One Dimensional Kac Equation}

\def\jem{Postal Address:
      Department of Applied Mathematics,
      University of Colorado, Box 526
      Boulder CO 80309-0526, USA; email: corcoran@colorado.edu, 
      phone: 303-492-0685}

\author {J. N. Corcoran, D. Jennings, and P. VaughanMiller\\
University of Colorado \thanks{\jem} }

\begin{document}


\maketitle
\vspace{-.8cm}

\begin{abstract} \small \noindent
We review the derivation of the Kac master equation model for random collisions of particles, its relationship to the Poisson process, and existing algorithms for simulating values from the marginal distribution of velocity for a single particle at any given time. We describe and implement a new algorithm that efficiently and more fully leverages properties of the Poisson process, show that it performs at least as well as existing methods, and give empirical evidence that it may perform better at capturing the tails of the single particle velocity distribution. Finally, we derive and implement a novel ``$\varepsilon$-perfect sampling'' algorithm for the limiting marginal distribution as time goes to infinity. In this case the importance is a proof of concept that has the potential to be expanded to more interesting (DSMC) direct simulation Monte Carlo applications.

\bigskip

\end{abstract}


\setcounter{page}{0}



\section{Introduction}
\slabel{int} 

In \cite{kac1956}, Kac introduced an $N$ particle stochastic process model for a spatially homogeneous gas that evolves according to a specific collision process. He proved that, as $N \rightarrow \infty$, the marginal probability density function (pdf) for a single particle at time $t$ approaches the solution of a simplified version of the famed Boltzmann equation. The $N$-dimensional probability density function for all particles is described by the {\bf{Kac master equation}} which can be derived as a random walk with transitions occurring at times of a Poisson process. In this paper we take advantage of properties of the Poisson process, including process ``thinning'' and the conditional uniformity of event times to efficiently simulate values from the velocity density of any single particle at any given time $t>0$. We compare our results to those from the familiar Nanbu, Nanbu-Babovsky, and DSMC (direct simulation Monte Carlo) algorithms. Additionally, we introduce an ``$\varepsilon$-perfect sampling algorithm'' that uses ``coupling from the past'' techniques to simulate or sample from the velocity distribution at time $t = \infty$ up to a user defined accuracy $\varepsilon$ that can be taken so small as to be essentially ``perfect'' within the accuracy available by a given machine precision.

In \Section{kacmaster}, we review the derivation of the $N$ particle Kac master equation and show its relationship to the Poisson process. In \Section{1dkac} we very briefly review the probability density function of any one given particle as a marginal density of the joint probability density function solution to the Kac master equation. In \Section{existing} we review three existing simulation methods for the marginal density, introduce the exact Poisson method, and make an empirical comparison between all methods. In \Section{new}, we briefly review the idea of perfect simulation (perfect sampling) and describe a novel ``$\varepsilon$-perfect'' algorithm that applies to the Kac model.


\section{The Kac Master Equation}
\slabel{kacmaster} 

Consider $N$ one-dimensional particles, each with mass $m$, with velocities $v_{1}, v_{2}, \ldots, v_{N}$. Define $\vec{v} = (v_{1}, v_{2}, \ldots, v_{N})$. The total kinetic energy of the ensemble of particles is
$$
E = \frac{m}{2} \sum_{i=1}^{N} v_{i}^{2}.
$$
For simplicity, we assume that $m=2$. Thus, we have $E = \sum_{i=1}^{N} v_{i}^{2}$  or, equivalently, $|\vec{v}| = \sqrt{E}$. 

If we assume that kinetic energy is conserved as these velocities evolve through particle collisions, points $\vec{v}$ with always be found on the sphere $S^{N-1}(\sqrt{E})$. 
We now describe a particular evolution of velocities, proposed by Kac \cite{kac1956} in two stages as in \cite{carlencarvahloloss} The first stage is to describe as a particular discrete time random walk  known as the ``Kac walk" and the second stage is to modify the random walk so that it is taking place in continuous time.

\subsection{The Kac Walk}

{\bf{The Kac Walk:}}

\begin{enumerate}
\item Randomly select a pair of indices $(i,j)$ from $\{1,2,\ldots,N\}$. There are $\left( \begin{array}{c}  N\\ 2 \end{array} \right)$ ways to do this, so a particular pair $(i,j)$ is selected with probability $\left[\left( \begin{array}{c}  N\\ 2 \end{array} \right)\right]^{-1}$

\item Randomly select a {\emph{scattering angle}} $\theta$ from a density $\rho(\theta)$ on $[0,2\pi)$. (Often $\theta$ is assumed to be uniformly distributed on $[0,2\pi)$. For an upcoming derivation, it will be convenient to define $\rho(\theta)$ on $[-\pi,\pi)$ and to assume that it is symmetric about $0$.)

\item Update the velocities for particles $i$ and $j$ through the rotation 
$$
(v_{i},v_{j}) \rightarrow (v_{i}^{\prime},v_{j}^{\prime}) := (v_{i} \cos \theta+v_{j} \sin \theta,-v_{i} \sin \theta + v_{j} \cos \theta).
$$
Return to Step 1.
\end{enumerate}

It is easy to see that total kinetic energy is conserved by the Kac walk (though momentum is not) and therefore that these three steps create a random walk on $S^{N-1}(\sqrt{E})$. We will now derive the probability density function, $f_{k}$ for the vector of $N$ velocities after $k$ ``collisions"  or iterations of the Kac walk.

Let $g:S^{N-1}(\sqrt{E}) \rightarrow \Re$ be any continuous function. Define an operator $Q_{N}$ on such functions as the conditional expectation
$$
Q_{N} g(\vec{v}) := \Expect[g(\vec{V}_{k+1})|\vec{V}_{k} = \vec{v}]
$$
where $\vec{V}_{k}$ is the random vector of velocities at the $k$th iteration of the Kac walk.

Conditioning on the selected angle and particles, this becomes
$$
\begin{array}{lcl}
Q_{N} g(\vec{v}) &=& \Expect[g(\vec{V}_{k+1})|\vec{V}_{k} = \vec{v}]\\
\\
&=& \displaystyle\int_{-\pi}^{\pi} \sum_{i<j} \Expect \left[ g(\vec{V}_{k+1}) \middle| \vec{V}_{k} = \vec{v}, \begin{array}{c} \mbox{\small particles $i$ and $j$}\\ \mbox{\small are selected}\end{array}, \begin{array}{c}  \mbox{\small angle $\theta$ is}\\ \mbox{\small selected} \end{array}     \right] \cdot \left[ \left(  \begin{array}{c} N \\ 2\end{array} \right) \right]^{-1} \, \rho (\theta) \, d \theta\\
\\
&=& \displaystyle\int_{-\pi}^{\pi} \sum_{i<j} g(v_{1}, \ldots, v_{i-1},v_{i}^{\prime}, v_{i+1}, \ldots, v_{j-1}, v_{j}^{\prime}, v_{j+1}, \ldots, v_{N}) 
\cdot \left[ \left(  \begin{array}{c} N \\ 2\end{array} \right) \right]^{-1} \, \rho (\theta) \, d \theta\\
\\
&=& \displaystyle\int_{-\pi}^{\pi} \sum_{i<j} g(R_{ij}(\theta) \, \vec{v}) 
\cdot \left[ \left(  \begin{array}{c} N \\ 2\end{array} \right) \right]^{-1} \, \rho (\theta) \, d \theta\\
\end{array}
$$ 
where  $R_{ij}(\theta)$ is the $N \times N$ rotation matrix that is the identity matrix altered to have  $\cos \theta$ as the $(i,i)$ entry, $\sin \theta$ as the $(i,j)$ entry, $-\sin \theta$ as the $(j,i)$ entry, and $\cos \theta$ as the $(j,j)$ entry.

Let $f_{0}$ be an initial probability density function for the $N$ velocities and let $f_{1}$ be the density after one iteration. Define random vectors $\vec{V}_{0} \sim f_{0}$ and $\vec{V}_{1} \sim f_{1}$. Note that
$$
\Expect[g(\vec{V}_{1})] = \int_{S} g(\vec{v}) \, f_{1}(\vec{v}) \, d \vec{v},
$$ 
where $S$ is shorthand for $S^{N-1}(\sqrt{E})$. Also note that
$$
\begin{array}{lcl}
\Expect[g(\vec{V}_{1})] &=& \Expect[\Expect[g(\vec{V}_{1})|\vec{V}_{0}]]\\
\\
&=& \int_{S} \Expect[g(\vec{V}_{1})|\vec{V}_{0}=\vec{v}] \cdot f_{0}(\vec{v}) \, d\vec{v}\\
\\
&=&  \int_{S} Q_{N} g(\vec{v}) f_{0}(\vec{v}) \, d\vec{v}
\end{array}
$$
Thus, we have that
\beq
\elabel{f0f1}
 \int_{S} g(\vec{v}) \, f_{1}(\vec{v}) \, d \vec{v} =  \int_{S} Q_{N} g(\vec{v}) f_{0}(\vec{v}) \, d\vec{v}.
\eeq

It is straightforward to verify that $Q_{N}$ is a self-adjoint operator in the sense that
\beq
\elabel{selfadjoint}
\int_{S} Q_{N} g(\vec{v}) f(\vec{v}) \, d\vec{v} = \int_{S} g(\vec{v}) Q_{N}(\vec{v}) f(\vec{v}) \, d\vec{v}
\eeq
for all continuous $g$ and for all $N$ dimensional densities $f$. Therefore, combining (\ref{e:f0f1}) and (\ref{e:selfadjoint}), we have
\beq
\elabel{f1isQf0}
\int_{S} g(\vec{v}) f_{1}(\vec{v}) \, d\vec{v} = \int_{S} g(\vec{v}) \, Q_{N} f_{0}(\vec{v}) \, d\vec{v}.
\eeq
Since (\ref{e:f1isQf0}) holds for all $g:S^{N-1}(\sqrt{E}) \rightarrow \Re$, we have that $f_{1} = Q_{N} f_{0}$. Similarly, the probability density function for the velocities after $k$ collisions is given by
\beq
\elabel{kcollpdf}
f_{k} = Q_{N}^{k} f_{0}.
\eeq

\subsection{Time and the Poisson Process}

As of now, we have a pdf for the $N$-dimensional vector of velocities evolving at the discrete times of a random walk. We now will allow the velocities to evolve as an $N$-dimensional continuous time Markov process. The pdf will now be denoted by $f(\vec{v},t)$. 

Kac proposed in \cite{kac1956} that collisions happen in the $N$-particle ensemble at times of a Poisson process. Specifically, it is assumed that the expected number of particles that collide in a vanishingly small time step $\Delta t$ is $\lambda N \Delta t$ for some $\lambda >0$. This means that the expected number of collisions between pairs of particles is $\lambda N \Delta t/2$  and so the assumption is that collisions happen according to a Poisson process with rate $\lambda N/2$.   Among other things, this implies that, as $\Delta t \searrow 0$, the probability of exactly one collision in any time interval $(t,t+\Delta t)$ is $\lambda N \Delta t/2 + o(\Delta t)$, the probability of no collisions is $1-\lambda N \Delta t/2 + o(\Delta t)$, and the probability of two or more is $o(\Delta t)$. (Here, $o(\Delta t)$ represents a vanishing function $g(\Delta t)$ such that $g(\Delta t)/\Delta t  \rightarrow 0$ as $\Delta t \searrow 0$.)

Given the velocity density $f(\vec{v},t)$, at time $t$, the density at time $t+\Delta t$ as $\Delta t \searrow 0$ is, informally, the  mixture density
$$
\begin{array}{lcl}
f(\vec{v},t+\Delta t) &=& p \cdot (\,\mbox{previous density after having gone through a collision}\,)\\
&&+ (1-p) \cdot (\,\mbox{previous density}\,)
\end{array}
$$
where $p$ is the probability of a collision in the ensemble during the time interval $(t,t+\Delta t)$ as $\Delta t \searrow 0$.

Formally, this is written as
$$
\begin{array}{lcl}
f(\vec{v},t+\Delta t) &=& (\lambda N \Delta t/2 + o(\Delta t)) \cdot Q_{N} f(\vec{v},t) + (1-\lambda N \Delta t/2 + o(\Delta t)) \cdot f(\vec{v},t)\\
\\
&=& \frac{\lambda N \Delta t}{2} \, Q_{N} f(\vec{v},t) + f(\vec{v},t) -\frac{\lambda N \Delta t}{2} \, f(\vec{v},t) +o(\Delta t)
\end{array}
$$
which implies that
$$
\frac{f(\vec{v},t+\Delta t)-f(\vec{v},t)}{\Delta t} = \frac{\lambda N [Q_{N}-I]}{2} f(\vec{v},t) + \frac{o(\Delta t)}{\Delta t}
$$
where $I$ is the identity operator.

Letting $\Delta t \searrow 0$, $\beta = \lambda/2$, and specifying an initial condition, we get the {\bf{Kac Master Equation}}:
$$
\begin{array}{lcl}
\frac{d}{dt} f(\vec{v},t) &=& \beta N [Q_{N}-I] f(\vec{v},t)\\
\\
f(\vec{v},0)&=& f_{0}(\vec{v})
\end{array}
$$
which has solution
$$
f(\vec{v},t) = e^{\beta N[Q_{N}-I]t} f_{0}(\vec{v}) = e^{-\beta N t} \sum_{i=0}^{\infty} \frac{(\beta N t)^{i}}{i!} Q_{N}^{i} f_{0}(\vec{v}).
$$


\section{The One Dimensional Kac Equation}
\slabel{1dkac}
We assume henceforth that the scattering angle density, $\rho(\theta)$, is the uniform density on the interval $(0,2\pi)$. We write $\theta \sim \unif (0,2 \pi)$.
Consider the marginal density for particle $1$,
\beq
\elabel{marginal}
f_{N}(v_{1},t) = \int_{S^{N-2}(\sqrt{E-v_{1}^{2}})} f(\vec{v},t) \, d\vec{v}_{-1}
\eeq
where $\vec{v}_{-1}:=(v_{2}, v_{3}, \ldots, v_{N})$. We will make the typical assumption that $E=N$ since the total kinetic energy is proportional to the number of particles in the system.

Kac \cite{kac1956} showed, under some assumptions\footnote{Let $\phi$ be a one dimensional density. Informally, an $n$-dimensional density $\phi_{n}$, symmetric in its arguments, is ``$\phi$-chaotic" if, as $n \rightarrow \infty$ any $k$-dimensional marginal of $\phi_{n}$ acts as a product of $k$ copies of $\phi$. In \cite{kac1956}, Kac proved that if $f^{(N)}=f(\vec{v},t)$ solves the Kac Master equation, then $f^{(N)}$ is $f$-chaotic where $f$ solves the Kac-Boltzmann equation. Thus, we can consider any univariate marginal of $f(\vec{v},t)$ as $N \rightarrow \infty$ as a solution to (\ref{e:onedimkac}). For a more formal explanation, please see \cite{carlenentropy}.} on the initial density $f(v,0)$,  that this single particle marginal density converges, as $N \rightarrow \infty$, to the solution of a one-dimensional spatially homogeneous Boltzmann-type equation:
\beq
\elabel{onedimkac}
\frac{df}{dt} (v,t) = \frac{\lambda}{2 \pi} \int_{-\infty}^{\infty} \int_{0}^{2\pi} [f(v^{\prime},t) f(w^{\prime},t)-f(v,t) f(w,t)] \, d \theta \, d w,
\eeq
where
$$
v^{\prime} = v \cos \theta + w \sin \theta \qquad \mbox{and} \qquad w^{\prime} = -v \sin \theta + w \cos \theta.
$$
Equation (\ref{e:onedimkac}) is known as the {\bf{Kac equation}} or the {\bf{Kac-Boltzmann equation}}.

Note that (\ref{e:onedimkac}) may be rewritten as
\beq
\elabel{kaqop}
\frac{\partial f}{\partial t} = \frac{1}{\varepsilon} \left[ P(f,f) - f \right]
\eeq
for $\varepsilon = 1/\lambda$, and $P(f,f)$ defined as the {\emph{bilinear collision operator}}
$$
P(f,f) = \frac{1}{2\pi} \int_{-\infty}^{\infty} \int_{0}^{2\pi} f(v^{\prime},t) f(w^{\prime},t) \, d \theta \, d w.
$$

If the initial distribution $f_{0}(v) :=f(v,0)$ on $(-\infty,\infty)$ is taken to be $f_{0}(v) = \frac{2}{\sqrt{\pi}} v^{2} e^{-v^{2}}$, and if $\lambda =\sqrt{\pi}/2$, the solution to (\ref{e:onedimkac})  is known \cite{krookandwu} to be
\beq
\elabel{onedimkacsol}
f(v,t) = \frac{1}{\sqrt{\pi}} \left[ \frac{3}{2} (1-C(t)) \sqrt{C(t)} + (3C(t)-1) (C(t))^{3/2} v^{2} \right] e^{-C(t) v^{2}}
\eeq
where
$$
C(t) = [3-2 \exp(-\sqrt{\pi} t/16)]^{-1}.
$$
It is easy to see that 
\beq
\elabel{onedimkacsolinf}
\lim_{t \rightarrow \infty} f(v,t) = \frac{1}{\sqrt{3 \pi}} \, e^{-\frac{1}{3} v^{2}} \qquad \mbox{for} \,\,\, -\infty<v<\infty.
\eeq

We will use (\ref{e:onedimkacsol}) and (\ref{e:onedimkacsolinf}) to test several Monte Carlo algorithms.


\section{Simulation Methods for $f(v,t)$}
\slabel{existing}

In Sections \ref{s:nanbu} and \ref{s:bird} we review existing methods for drawing (simulating) velocity values from the single particle marginal distribution $f_{N}(v,t)$ for finite $t$. In \Section{poisson} we propose a new method that takes advantage of properties of the Poisson collision process. Our goal here is to have a method  efficient enough to be able to push $N$ quite large so that we may see approximately $f(v,t)$. 

\subsection{The Nanbu and Nanbu-Babovsky Algorithms}
\slabel{nanbu}
In \cite{nanbu} Nanbu proposes an algorithm to draw from $f_{N}(v,t)$ based on the probabilistic interpretation of colliding particles and on the discretization
\beq
\elabel{discrete}
f_{n+1} = \left( 1- \lambda \Delta t \right) f_{n} + \lambda \Delta t P(f_{n},f_{n})
\eeq
of (\ref{e:kaqop}). Here, $f_{n} = f_{n}(v) = f(v,t_{n})$, $t_{0}=0$, and $t_{n} = t_{0}+n \Delta t$ for some small increment $\Delta t>0$. The interpretation of (\ref{e:discrete}) is that, if $\Delta t$ is small enough so that $0 \leq \lambda \Delta t \leq 1$, ``particle 1" (or any fixed given particle) 
\begin{itemize} 
\item fails to undergo a collision and does not have any velocity update with probability $1-\lambda \Delta t$, or
\item experiences a collision with probability $\lambda \Delta t$ and has a new velocity generated according to the collision process described by $P(f,f)$.
\end{itemize}

For the following algorithm descriptions, we assume that $\Delta t$ divides into $t$ evenly.

\vspace{0.2in}
{\bf{Nanbu's Algorithm}}

For $i=1,2,\ldots, N$, let $v_{i}^{(n)}$ denote the velocity for particle $i$ at time step  $n=0,1,2,\ldots$.
\begin{enumerate}
\item Sample initial velocities, $v_{1}^{(0)}, v_{2}^{(0)}, \ldots, v_{N}^{(0)}$, for all $N$ particles independent and identically distributed (iid) from $f_{0}(v)$. Set $s=0$.
\item For $i=1,2,\ldots, N$,
\begin{itemize}
\item With probability $1-\lambda \Delta t$ do nothing to the velocity of the $i$th particle. 

Set $v_{i}^{(n+1)} = v_{i}^{(n)}$.
\item With probability $\lambda \Delta t$,
\begin{itemize}
\item[$\circ$] Randomly select a second particle from among the remaining particles in $\{1,2,\ldots, N\}$.
\item[$\circ$] Draw an angle value $\theta \sim \rho(\theta)$.
\item[$\circ$] Set $v_{i}^{(n+1)} = v_{i}^{(n)} \cos \theta + v_{j}^{(n)} \sin \theta$
\end{itemize}
\end{itemize}
\item Set $s=s+\Delta t$ and $n=n+1$. If $s < t$, return to Step 2. 
\end{enumerate}

\vspace{0.2in}
As kinetic energy is not conserved in Nanbu's Algorithm, Babovsky \cite{babovsky} modified the algorithm to update both $v_{i}$ and $v_{j}$ as particles $i$ and $j$ collide. Additionally, at each time step, the number of collisions processed is taken to be equal to the expected number of collision pairs in an interval of length $\Delta t$. As the expected number of collisions is $\lambda N \Delta t$, the expected number of collision pairs is $\lambda N \Delta t/2$.  This  is then  probabilistically rounded using the following function.
$$
Round[x] = \left\{
\begin{array}{lcl}
\lfloor x \rfloor &,& \mbox{with probability} \,\,\, \lfloor x \rfloor +1-x\\
\lfloor x \rfloor +1 &,& \mbox{with probability} \,\,\, x-\lfloor x \rfloor\\
\end{array}
\right.
$$
It is easy to see that, for any given $x$, the  expected value of $Round[x]$ is $x$.

\vspace{0.2in}
{\bf{Nanbu-Babovsky Algorithm}}

For $i=1,2,\ldots, N$, let $v_{i}^{(n)}$ denote the velocity for particle $i$ at time step  $n=0,1,2,\ldots$.
\begin{enumerate}
\item Sample initial velocities, $v_{1}^{(0)}, v_{2}^{(0)}, \ldots, v_{N}^{(0)}$,  for all $N$ particles independent and identically distributed (iid) from $f_{0}(v)$. Set $s=0$.
\item For $i=1,2,\ldots, N$,
\begin{itemize}
\item Select $m=Round[\lambda N \Delta t /2]$.
\item Randomly select $m$ pairs, $(i,j)$ of particle indices, without replacement, from among all possible pairs.
\item For each selected pair $(i,j)$,  draw an angle value $\theta \sim \rho(\theta)$ and set
$$
\begin{array}{lcl}
v_{i}^{(n+1)} &=& v_{i}^{(n)} \cos \theta + v_{j}^{(n)} \sin \theta\\
v_{j}^{(n+1)} &=& -v_{i}^{(n)} \sin \theta + v_{j}^{(n)} \cos \theta.
\end{array}
$$
\item For the remaining unselected indices $k$, set $v_{k}^{(n+1)}=v_{k}^{(n)}$.
\end{itemize}
\item Set $s=s+\Delta t$ and $n=n+1$. If $s < t$, return to Step 2. 
\end{enumerate}

Both the Nanbu and the Nanbu-Babovsky algorithms can be shown to converge to the solution of the discretized Kac equation given by (\ref{e:discrete}). Nanbu's algorithms allows each particle to potentially undergo a collision only once per time step and the Nanbu-Babovsky scheme allows each pair of particles to potentially undergo a collision once per time step.  

\subsection{Bird's DSMC Algorithm}
\slabel{bird}
The Direct Simulation Monte Carlo (DSMC) approach due to Bird \cite{bird94} was first proposed in the late 60's and has been shown \cite{wagner} to converge to the solution of (\ref{e:kaqop}). It differs from the Nanbu-Babovsky algorithm in that any particular collision pair may be selected multiple times in the time interval of length $\Delta t$. Since there are $\lambda N \Delta t /2$ collisions expected in that interval, the average time between collisions is taken to be $\Delta t_{c} := \Delta t / [\lambda N \Delta t /2]  = 2 / (\lambda N)$.

{\bf{Bird's DSMC Algorithm}}

\begin{enumerate}
\item Sample initial velocities, $v_{1}^{(0)}, v_{2}^{(0)}, \ldots, v_{N}^{(0)}$,  for all $N$ particles independent and identically distributed (iid) from $f_{0}(v)$. Set $s=0$. 

\item Process one collision as follows.

\begin{itemize}
\item Select a pair $(i,j)$ of particles at random from among all possible pairs.
\item For each selected pair $(i,j)$,  draw an angle value $\theta \sim \rho(\theta)$ and set
$$
\begin{array}{lcl}
v_{i}^{(n+1)} &=& v_{i}^{(n)} \cos \theta + v_{j}^{(n)} \sin \theta\\
v_{j}^{(n+1)} &=& -v_{i}^{(n)} \sin \theta + v_{j}^{(n)} \cos \theta.
\end{array}
$$
\item For the remaining unselected indices $k$, set $v_{k}^{(n+1)}=v_{k}^{(n)}$.

\end{itemize}

\item Set $s = s + \Delta t_{c}$ and $n=n+1$.  If $s<t$, return to Step 2.

\end{enumerate}

Bird's DSMC techniques comprise a much wider class of extremely useful and popular algorithms than the one detailed here. As it stands, the one given here is essentially the Nanbu-Babovsky algorithm where pairs are selected with replacement.

\subsection{Exact Simulation of $f_{N}(v,t)$ Based on Properties of the Poisson process}
\slabel{poisson}

In this section, we simulate exactly from $f_{N}(v,t)$ using the actual Poisson numbers of collisions between important time points as opposed to the expected numbers of collisions. We do not partition the interval $(0,t)$ as in the previously described algorithms but instead focus on collision times for the fixed ``particle 1''.

Recall that the $N$-particle ensemble is assumed to undergo collisions at times of a Poisson process with rate $\lambda N$. As each collision involves 2 particles, this means that pairs of particles are colliding according to a Poisson process with rate $\lambda N /2$. The probability that particle $1$ is involved in any one of these collisions is $(N-1) / {N \choose 2}=2/N$. So, the particle $1$ collision process is a thinned Poisson process with rate $(\lambda N/2) \cdot 2/N = \lambda$. Therefore, we begin by simulating a Poisson rate $\lambda t$ number of collisions for particle $1$ in the time interval $(0,t)$. 

Though intercollision times for particle $1$ are exponential with rate $\lambda t$, once the number of collisions during $(0,t)$ is fixed, the times of those collisions are uniformly distributed over the interval $(0,t)$. So, rather than divide up the interval $(0,t)$ into $\Delta t$ increments, we consider only the times when particle $1$ is involved in a collision by first simulating $K \sim Poisson (\lambda t)$, then simulating independent $U_{1}, U_{2}, \ldots, U_{K}$ from the uniform distribution over $(0,t)$ and finally considering the ordered time points $T_{1}, T_{2}, \ldots, T_{K}$ corresponding to the ordered $U_{i}$. (For example, $T_{1} = \min (U_{1}, U_{2}, \ldots, U_{K})$ and $T_{K} = \max (U_{1}, U_{2}, \ldots, U_{K})$.)

To update the velocity for particle $1$ at time $T_{1}$, we must choose, at random, a particle from the remaining $N-1$ particles for collision. As this ``sub-ensemble'' of particles has been evolving over the time interval $(0,T_{1})$, we will process $K_{1} \sim Poisson ( \lambda (N-1) T_{1}/2)$ collisions involving only these particles without any regard for the times of these collisions and then select one at random for collision with particle $1$. Continuing in this manner, we get the following algorithm.

{\bf{Exact Poisson Algorithm}}

\begin{enumerate}
\item Sample initial velocities, $v_{1}^{(0)}, v_{2}^{(0)}, \ldots, v_{N}^{(0)}$, for all $N$ particles independent and identically distributed (iid) from $f_{0}(v)$. 

\item Simulate $K \sim Poisson (\lambda t)$. This is the number of collisions involving particle $1$ during $[0,t]$. Suppose that $K=k$.

\item Simulate $U_{1}, U_{2}, \ldots, U_{k} \stackrel{iid}{\sim} \unif (0,t)$ collision times for particle $1$. Denote the ordered $U_{1}, U_{2}, \ldots, U_{k}$ as $T_{1}, T_{2}, \ldots, T_{k}$ where $T_{1} = \min(U_{1}, U_{2}, \ldots, U_{k})$ and $T_{k} = \max (U_{1}, U_{2}, \ldots, U_{k})$.

\item Let $T_{0}=0$. 

For $i=1,2, \ldots, k$,
\begin{itemize}
\item Simulate $K_{i} \sim Poisson (\lambda (N-1)(T_{i}-T_{i-1})/2)$. This is the number of ``non particle 1'' collisions happening during the time interval $[T_{i-1},T_{i}]$. Suppose that $K_{i}=k_{i}$.
\item Process $k_{i}$ collisions for the $(N-1)$ particle ensemble as follows.

For $j=1,2, \ldots, k_{i}$
\begin{itemize}
\item[$\circ$] Select a pair $(r,s)$ of particles at random from among all possible pairs or particles with labels in $\{2,3, \ldots, N\}$.

\item[$\circ$] For each selected pair $(r,s)$,  draw an angle value $\theta \sim \rho(\theta)$ and set new velocities as
$$
\begin{array}{lcl}
v_{r} &=& v_{r} \cos \theta + v_{s} \sin \theta\\
v_{s} &=& -v_{r} \sin \theta + v_{s} \cos \theta
\end{array}
$$ 
where $v_{r}$ and $v_{s}$ on the right-hand side of these equations are the pre-collision velocities for particles $r$ and $s$.
\end{itemize}
\item Randomly select one particle from $\{2,3, \ldots, N\}$ and process a collision with particle $1$ by assigning new velocities
$$
\begin{array}{lcl}
v_{r} &=& v_{r} \cos \theta + v_{s} \sin \theta\\
v_{s} &=& -v_{r} \sin \theta + v_{s} \cos \theta
\end{array}
$$ 
where $r$ is the selected index in $\{2,3,\ldots,N\}$ and  $v_{1}$ and $v_{r}$ on the right-hand side of these equations are the pre-collision velocities for particles $1$ and $r$. 
\end{itemize}

\end{enumerate}

Note that, unlike the Nanbu, Nanbu-Babovsky, and Bird algorithms, we are not using any sort of time counter or partitioning of the time interval $[0,t]$. Furthermore, we are actually processing a Poisson number of collisions, as assumed by the derivation of the Kac equation, in any given time interval rather than simply the expected number of collisions under the Poisson assumption as in Bird's algorithm.  Finally, we note that we are saving on collisions by not propagating the system in the time interval $[T_{K},t]$ as we are trying to sample the velocity of particle $1$ at time $t$ and this is unchanged in this interval. This gives an expected savings of  $(N/2)(1-e^{-\lambda t})$ collisions per simulation repetition.

\subsection{Simulation Results}

In order to assess the performance of the different algorithms, we simulated repeated draws from the velocity distribution of the first particle, $v_{1}$, for various values of $N$, and compared the results to the exact solution (for $N \rightarrow \infty$) given by (\ref{e:onedimkacsol}). Specifically, for each algorithm we simulated $N$ iid velocities and propagated the ensemble forward according to the algorithms described in Sections \ref{s:nanbu}, \ref{s:bird}, and \ref{s:poisson}, collecting the ultimate values for $v_{1}$ at various times $t$. In all cases, we were able to achieve reasonable accuracy for a surprisingly small number of particles. However, since the accuracy of the Nanbu and Nanbu-Babovsky algorithms both depend on how finely the time interval is partitioned and since the Nanbu-Babovsky sampling with replacement step is relatively inefficient, we will focus our comparisons mainly on the Nanbu, Bird, and Poisson algorithms.

We will express the accuracy of our numerical results using an estimate of the {\bf{total variation norm}} ($TVN$) distance between the true curve and the estimated distributions. The total variation norm distance between two probability measures $\pi_{1}$ and $\pi_{2}$ is defined to be
$$
||\pi_{1}-\pi_{2}|| = \frac{1}{2} \sup_{A \in {\cal {B}}} |\pi_{1}(A)-\pi_{2}(A)|,
$$
where ${\cal {B}}$ is the set of Borel sets in the appropriate space and $\pi_{i}(A) = \int_{A} \pi_{i}(x) \, dx$. If $\pi_{1}$ and $\pi_{2}$ are discrete measures, then the total variation norm can be written simply as
\beq
\elabel{discretetvn}
||\pi_{1}-\pi_{2}|| = \frac{1}{2} \sum_{x} |\pi_{1}(x)-\pi_{2}(x)|.
\eeq 
In this paper, we discretized both the true and simulated densities by a natural histogram binning procedure in order to use (\ref{e:discretetvn}) to give an estimated $TVN$ distance between the target density and our simulation results. As this estimate depends on the bin widths used, the actual values of our estimated $TVN$ are not as interesting as the relative values. We will keep our bin widths constant throughout.

\begin{figure}
\caption{Histograms of 100,000 values of $v_{1}$ at $t=2$ with $N=5$ for $\Delta t = 0.01$ (left) and $\Delta t = 1.0$ (right). Curve is exact solution for $N= \infty$ given by (\ref{e:onedimkacsol}).}
\label{fig:nanbu01}
\begin{subfigure}{0.5\textwidth}
\includegraphics[width=\linewidth,height=2.2in]{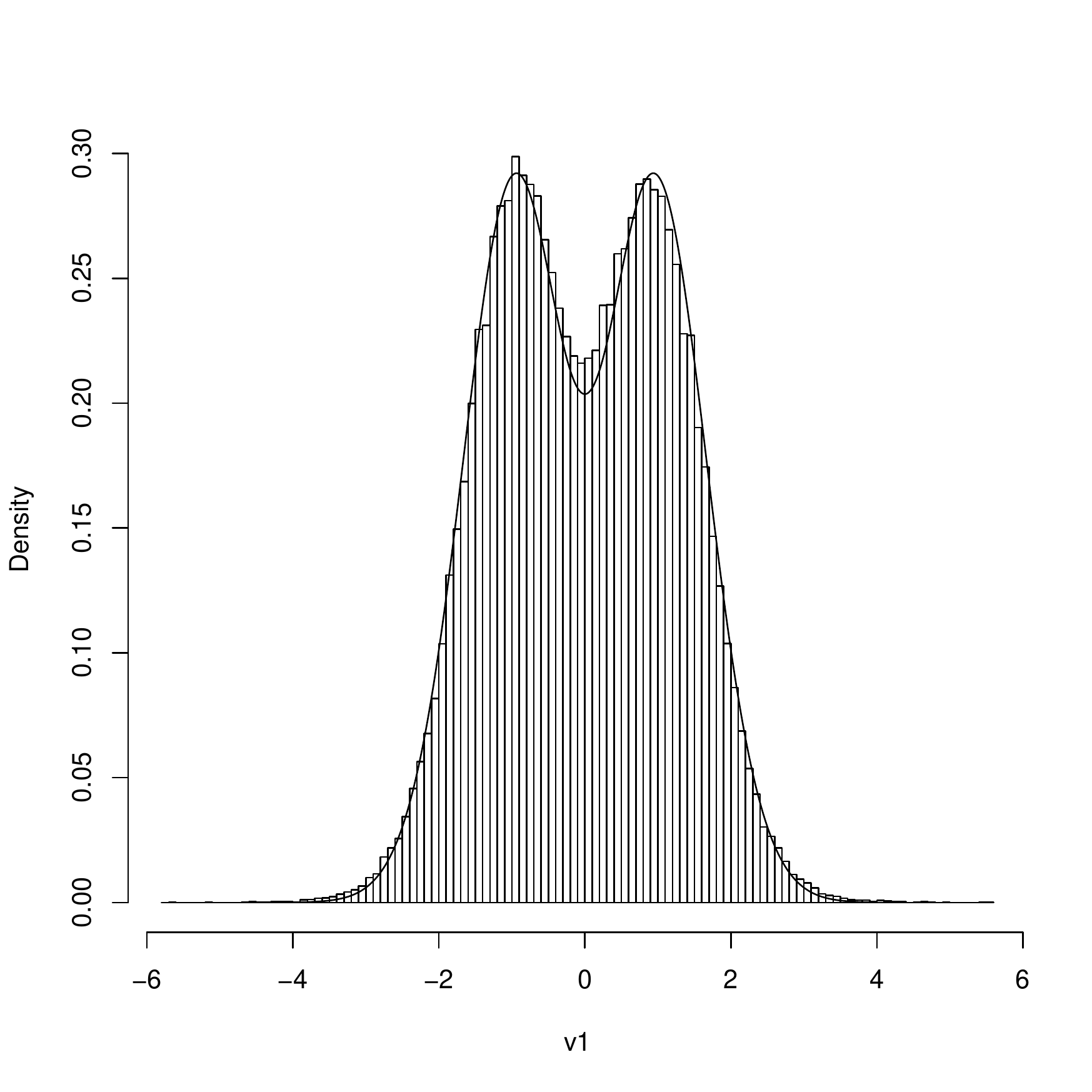}
\end{subfigure}
\hspace*{\fill} 
\begin{subfigure}{0.5\textwidth}
\includegraphics[width=\linewidth,height=2.2in]{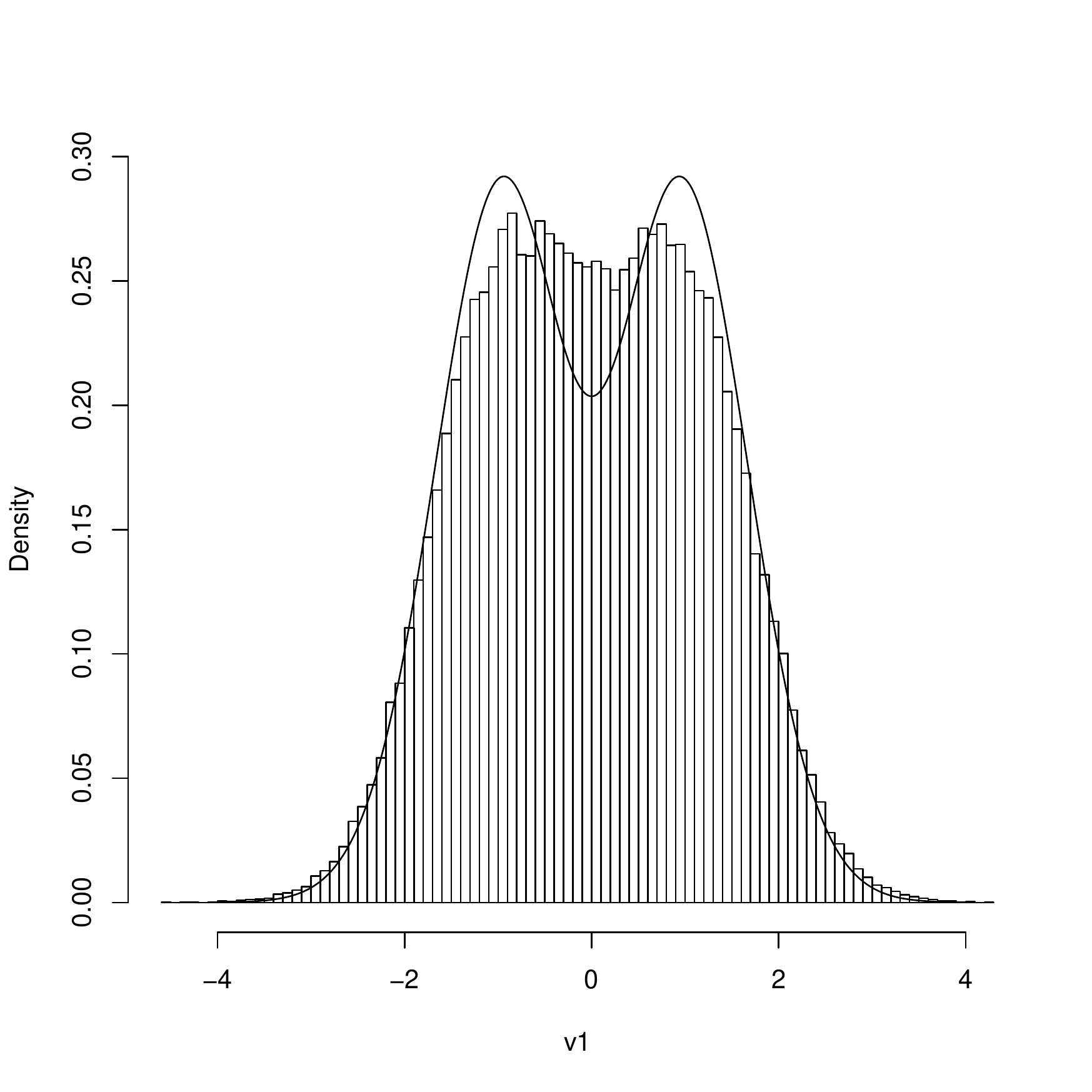}
\label{fig:shb}
\end{subfigure}
\end{figure}

\begin{figure}
\caption{Estimated $TVN$ for Nanbu's Algorithm as a function of $k$ for $\Delta t = 2/2^{k}$}
\label{fig:nanbuTVN}
\begin{center}
\includegraphics[width=4in,height=3.3in]{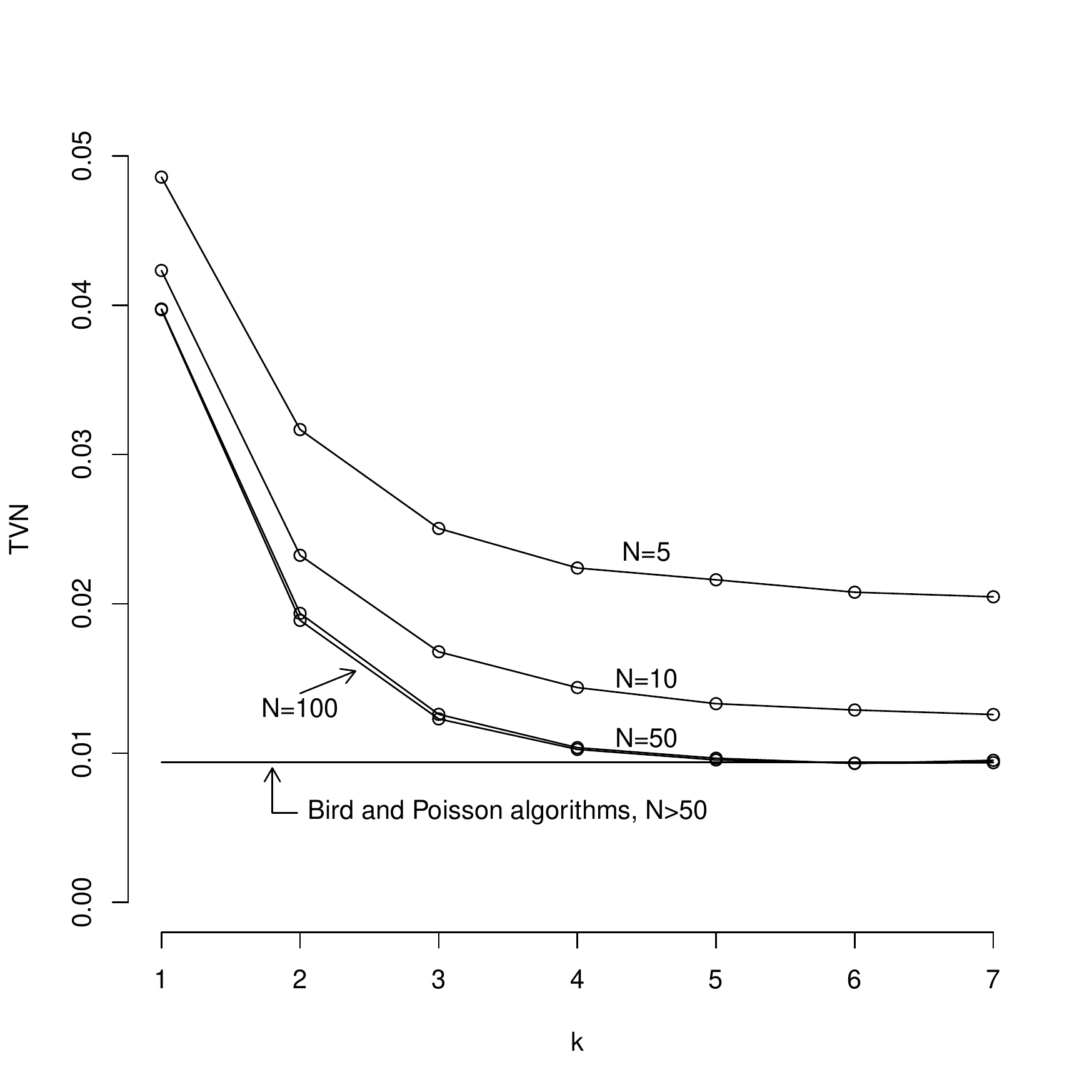}
\end{center}
\end{figure}

\begin{figure}
\caption{Estimated $TVN$ for the Nanbu, Bird, and Poisson algorithms as a function of $N$}
\label{fig:compareTVN}
\begin{center}
\includegraphics[width=4in,height=3.3in]{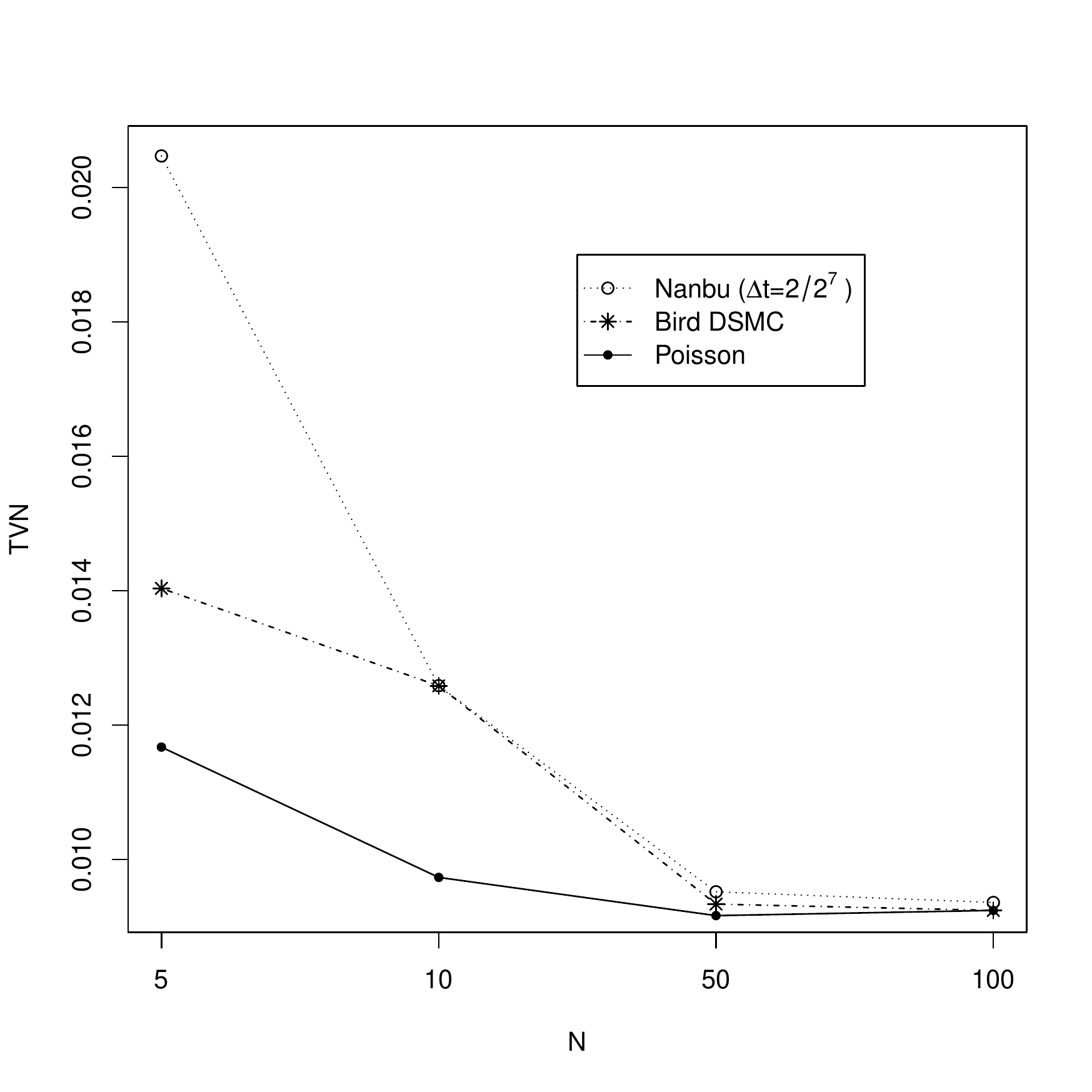}
\end{center}
\end{figure}

\begin{figure}
\caption{Histograms of 100,000 Values of $v_{1}$ at $t=2$ with $N=50$ for the DSMC algorithm (left) and the Poisson algorithm (right).} 
\label{fig:birdpoisshist}
\begin{subfigure}{0.5\textwidth}
\includegraphics[width=\linewidth,height=2.2in]{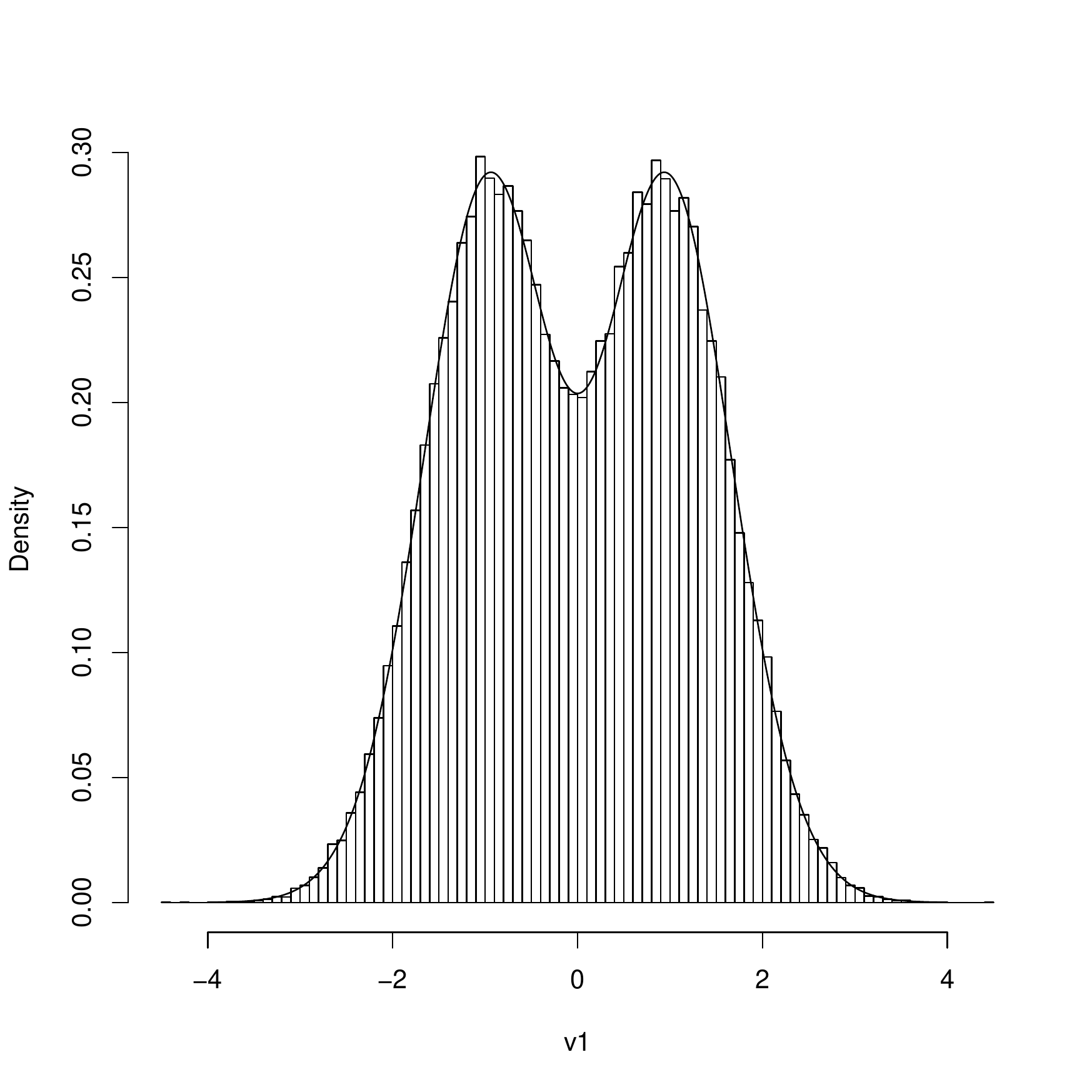}
\end{subfigure}
\hspace*{\fill} 
\begin{subfigure}{0.5\textwidth}
\includegraphics[width=\linewidth,height=2.2in]{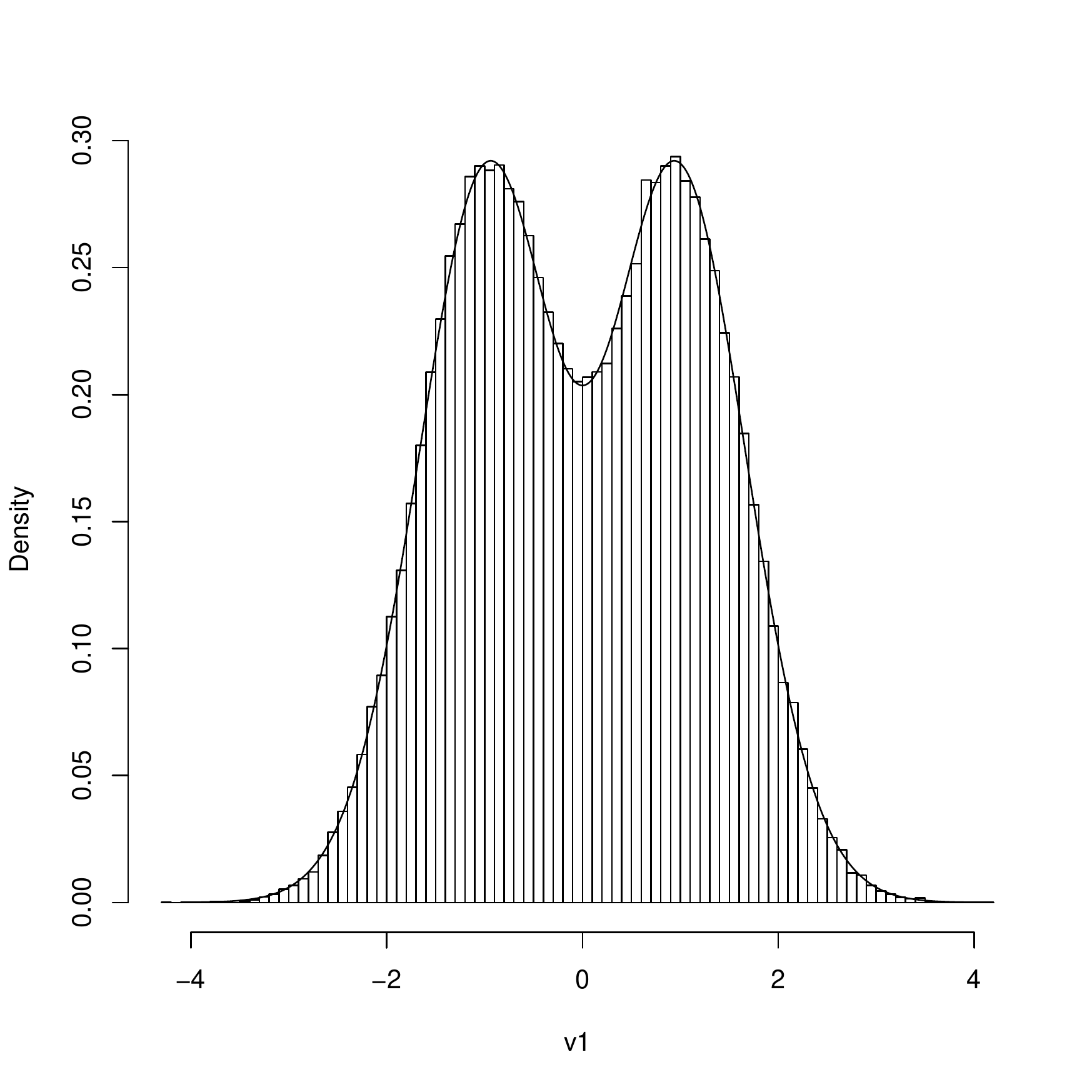}
\label{fig:shb}
\end{subfigure}
\end{figure}

\begin{figure}
\caption{Histograms of 100,000 upper tail values of $v_{1}$ at $t=2$ with $N=1000$ using Bird's DSMC algorithm (left) and the Poisson algorithm (right).}
\label{fig:tails}
\begin{subfigure}{0.5\textwidth}
\includegraphics[width=\linewidth,height=2.2in]{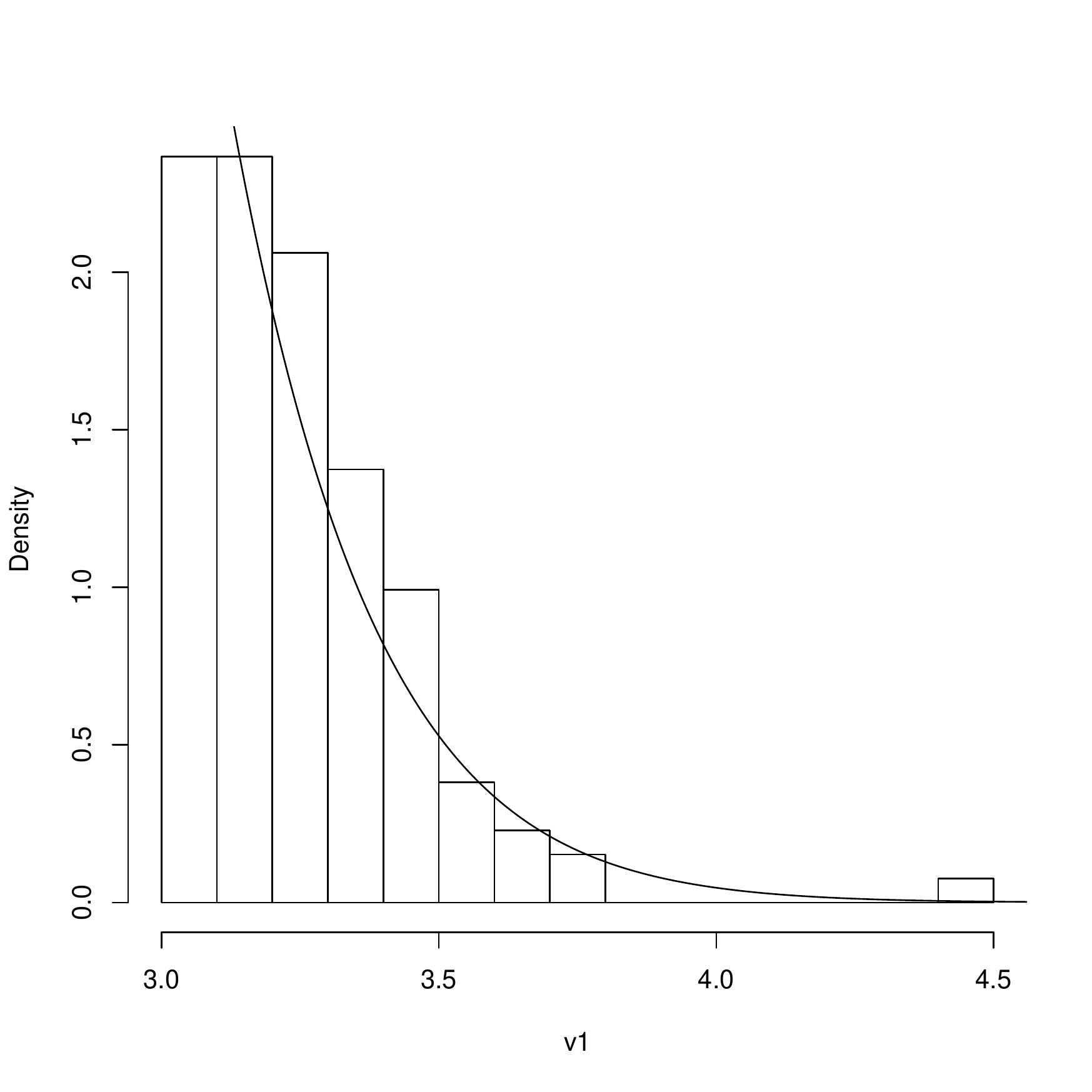}
\end{subfigure}
\hspace*{\fill} 
\begin{subfigure}{0.5\textwidth}
\includegraphics[width=\linewidth,height=2.2in]{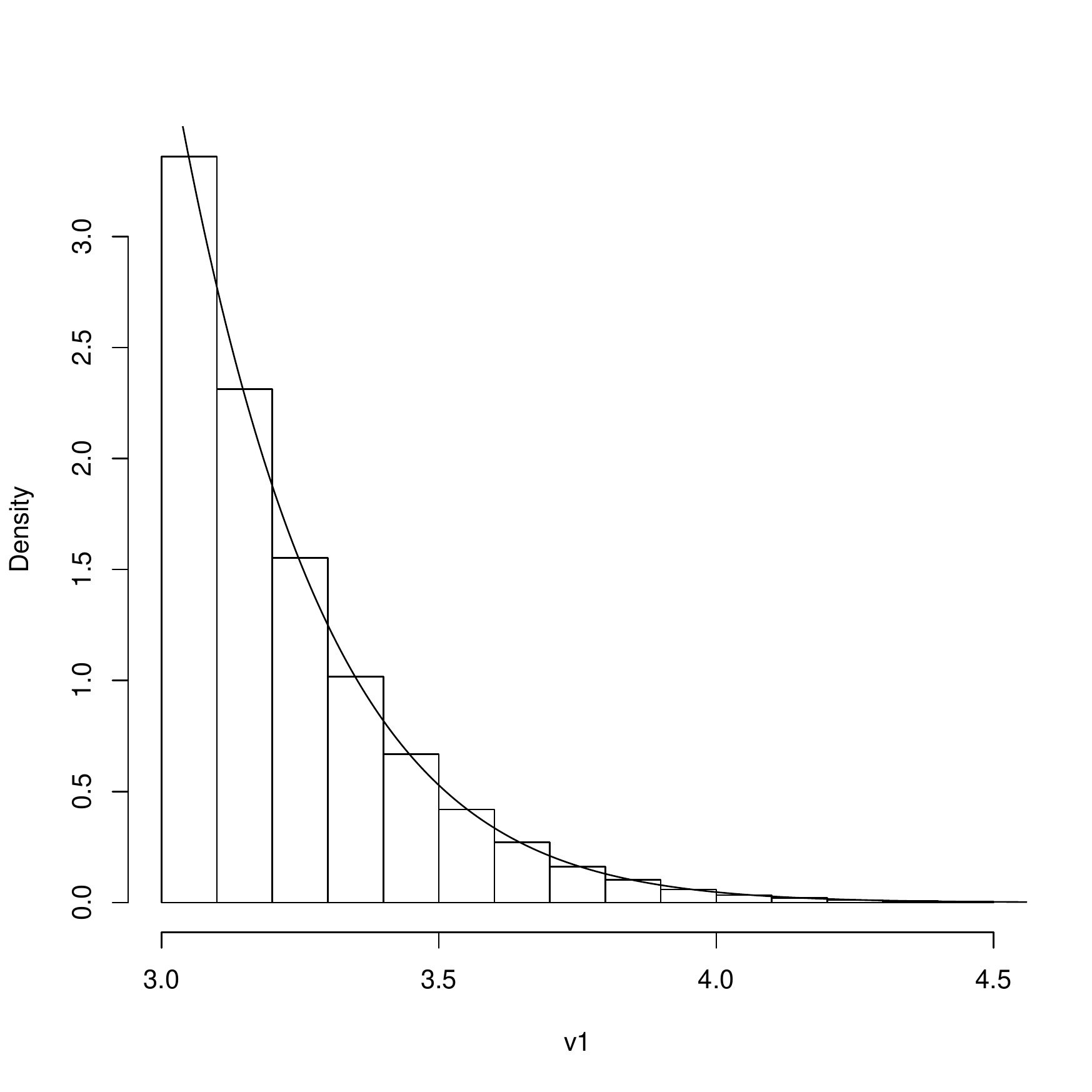}
\label{fig:shb}
\end{subfigure}
\end{figure}

For example, for $t=2$, using Nanbu's algorithm with $N=5$ produced distributions for $v_{1}$ shown in Figure \ref{fig:nanbu01}. In both cases the ``target density'' given by (\ref{e:onedimkacsol}) is shown as a solid curve.
It is no surprise that the smaller simulation time step gives better results, but to illustrate the $TVN$ error, the estimated values of $TVN$ for the left and right graphs are $0.0187$ and $0.0482$, respectively. Figure \ref{fig:nanbuTVN} shows estimated $TVN$, for several values of $N$, as  the simulation time step size decreases. Each point on the graph was created by averaging $100$ values of $TVN$, each having been computed from a sample of size $100,000$. The height of the horizontal line just below $0.01$ shows the average $TVN$ for $100$ samples of size $100,000$ for Bird's DSMC algorithm (which doesn't depend on $\Delta t$) for $N=50$. This was indistinguishable to several decimal places from similarly estimated $TVN$ for larger $N$, for the Poisson algorithm for $N \geq 50$, and for a baseline estimated $TVN$ for samples of size $100,000$ from the standard normal distribution. (Recall that we do not expect the estimated $TVN$ to necessarily approach zero even in the case of exact simulation from the normal distribution due to the discretization used in its calculation.) Figure \ref{fig:compareTVN} shows the average $TVN$ values (again averaging $100$ values each computed from a sample of size $100,000$) for the three algorithms as a function of the number of particles in the simulations. Here we can see that the Poisson algorithm consistently performs better than both the Nanbu and Bird algorithms for small $N$. That is, less particles are needed in order to approximate (\ref{e:onedimkacsol}) when using the Poisson algorithm.  Distributions for $v_{1}$, estimated by the DSMC and Poisson algorithms are shown in Figure \ref{fig:birdpoisshist}. While they are similar both visually and by the $TVN$ measure, the Poisson algorithm appears to be performing slightly better in the tails, as shown in Figure \ref{fig:tails} for the case of the upper tails.  

\section{An $\varepsilon$-Perfect Simulation of $f(v,\infty)$}
\slabel{new}

``Perfect sampling'' (or ``perfect simulation'') is the name applied to a class of Markov chain Monte Carlo (MCMC) algorithms which enable  one to draw values exactly from the
stationary (limiting) distribution $\pi$ of a Markov chain. This is in contrast to the more typical MCMC algorithms in which transitions are simulated for ``a long time'' after which output is collected giving only approximate draws from $\pi$.

\subsection{Perfect Simulation}
\slabel{ps}
The essential idea of most of these approaches is to find a random epoch
$-T$ in the past such that, if one constructs sample paths (according to a
transition density $p(x,y)$ that has stationary/limiting distribution $\pi$) from every point in
the state space starting at $-T$,  all paths will have met or ``coupled''
by time zero. The common value of the paths at time zero is a
draw from $\pi$.
Intuitively, it is clear why  this result holds with such a
random time $T$.
Consider a chain starting anywhere in the state space  at time $-\infty$. 
At time $-T$ it must pick
{\it some} value $x$, and from then on it follows the trajectory from that
value. Since
it arrives at
the same place at time zero no matter what value $x$ is picked at time $-T$,
the value returned by the algorithm at time zero is the tail end of a path that has run for an infinitely long time and is therefore a ``perfect'' draw
from $\pi$. 

\begin{figure}
\caption{Depiction of stochastically dominating (bounding) processes in perfect simulation.}
\label{fig:sandwich}
\centering
\includegraphics[width=4in,height=3.5in]{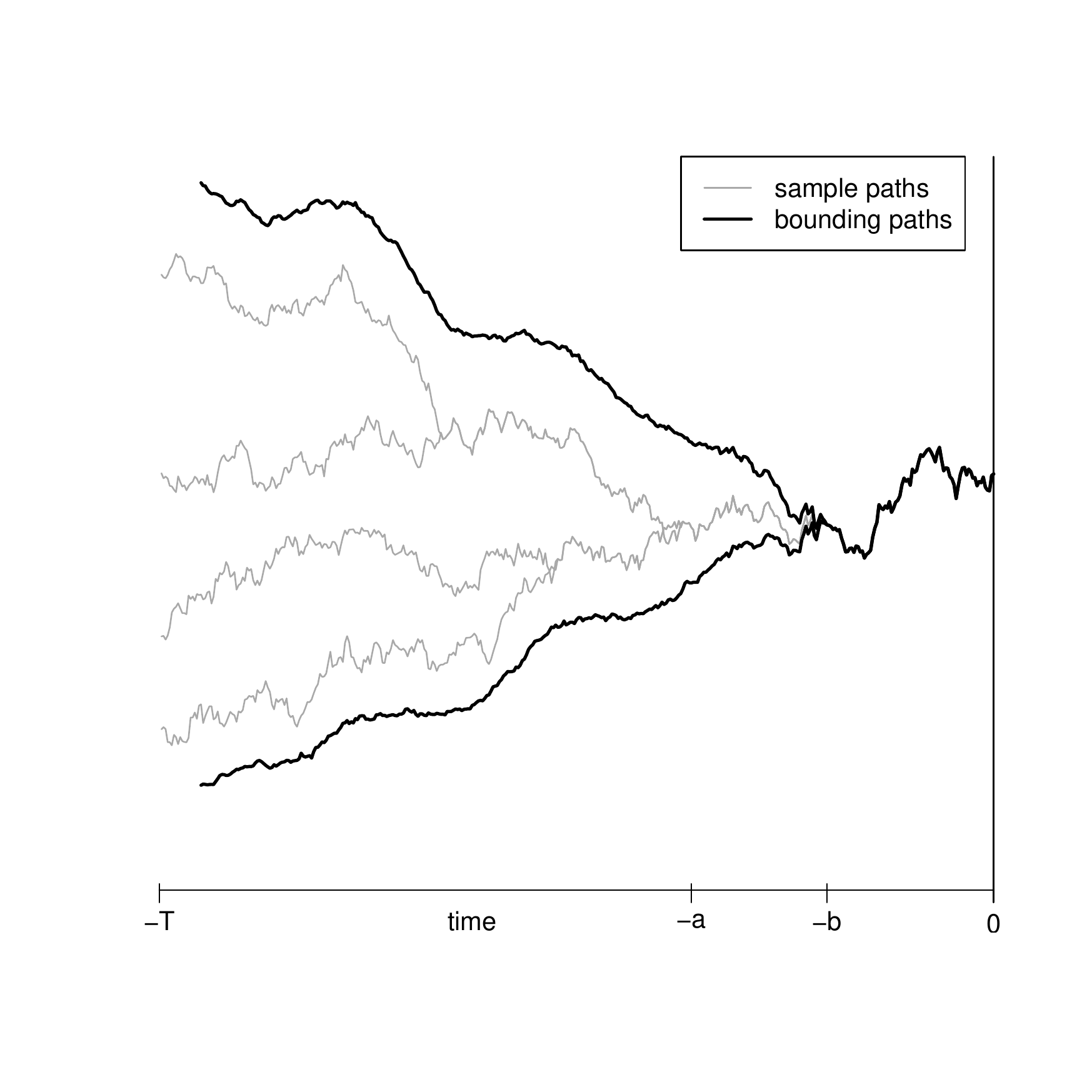}
\end{figure}

Typically, perfect sampling is carried out using {\emph{stochastically dominating processes}} that bound or ``sandwich'' the process of interest for all time as depicted in Figure \ref{fig:sandwich}. At time $-T$, these bounding processes are at the extremes of the state space. At time $-b$, they have coupled together to form a single path that is followed forward to time $0$. If such coupling is not achieved before time $0$, $T$ is increased and the bounding processes are resimulated using the same random inputs whenever they pass through previously visited time points. Note that when the bounding processes couple at some time $-b$, all possible sample paths of the process of interest have necessarily coupled at some earlier time $-a$. Note further that a path started anywhere in the state space at time $-T$ will be at the same place at time $0$. This value at time $0$ is a perfect draw from the stationary distribution of the Markov chain. Time $-T$ is known as a {\emph{backward coupling time}}.

Forward coupling of sample paths does not yield the same result for reasons that we will not go into here. For more information on perfect sampling algorithms, we refer the reader to \cite{casrob00}.

\subsection{$\varepsilon$-Perfect Simulation for the Kac Model}
\slabel{epsperf}

We say that an algorithm is ``$\varepsilon$-perfect'' if the distance between the bounding processes depicted in Figure \ref{fig:sandwich} is monotonically decreasing and if this distance is less than $\varepsilon$ at time $0$. In this Section we will describe an $\varepsilon$-perfect algorithm for sampling from $f(v,\infty)$ that is interesting mainly as a proof of concept for the approach that we hope will eventually be adapted to and useful for more interesting  models. 

Consider, for the purpose of illustration, the case where we have only $N=3$ particles. Simulating iid velocities $v_{1}$, $v_{2}$, and $v_{3}$ from $f_{0}(v)$ gives a kinetic energy $E = v_{1}^{2}+v_{2}^{2}+v_{3}^{2}$. As we propagate these $3$ velocities forward in time, we create a realization of a 3-dimensional Markov chain that lives on the surface of a 3-dimensional sphere with radius $\sqrt{E}$. For the simple model considered in this paper, we can, without loss of generality, consider the velocity vector as a random walk on the sphere restricted to the first octant where $v_{1},v_{2},v_{3}>0$ and at the end of our simulation attach a negative sign to each velocity independently of the others with probability $1/2$.

\begin{figure}
\caption{Visualization of sample paths of the ``Perfect Kac Walk'' on a sphere in the first octant.}
\label{fig:spherepoints}
\begin{center}
\includegraphics[width=2.8in,height=2.6in]{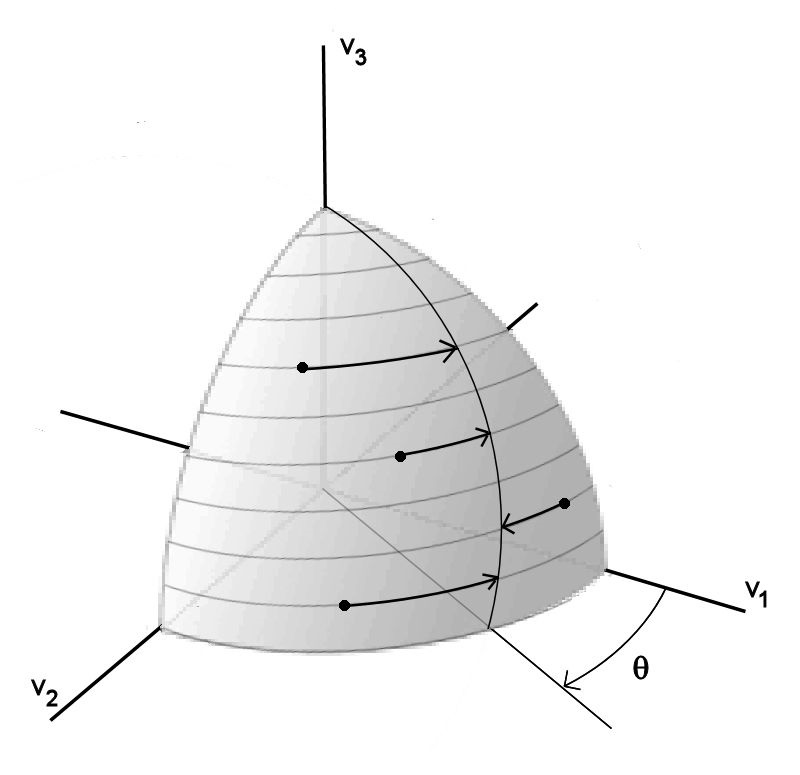}
\end{center}
\end{figure}

For the purpose of achieving a coupling of sample paths walking on the sphere, we note that $v_{i} \cos \theta + v_{j} \sin \theta$ can be written as
$$
v_{i} \cos \theta + v_{j} \sin \theta = A \, \sin (\theta + \varphi)
$$
for some $A$ and $\varphi$. In the case where $v_{j}=0$, we may take $A=v_{i}$ and $\varphi = \pi/2$. In the case that $v_{j} \ne 0$, we may take $A = \sqrt{v_{i}^{2}+v_{j}^{2}}$ and $\varphi = \tan^{-1} (v_{i}/v_{j})$. Suppose that the current position of the random walk is $(v_{1}, v_{2}, v_{3})$ and that the particles randomly selected for updating are indexed by $(i,j)=(1,2)$. Note then that $A = \sqrt{E-v_{3}^{2}}$ and, since $v_{2}$ is non-zero with probability $1$, the new velocity for particle $1$ is given by
\beq
\elabel{v1sine}
v_{1}^{\prime} = \sqrt{E-v_{3}^{2}} \,\, \sin (\theta + \varphi)
\eeq
where $\theta \sim \unif(0,2 \pi)$ and $\varphi = \tan^{-1} (v_{1}/v_{2})$. Since $\theta+\varphi$ is uniformly distributed over an interval of length $2 \pi$, the distribution of $v_{1}^{\prime}$ in (\ref{e:v1sine}) is the same as that of $\sqrt{E-v_{3}^{2}} \,\, \sin (\theta)$. Thus, we will update $v_{1}$ as  
\beq
\elabel{v1sine2}
v_{1}^{\prime} = \sqrt{E-v_{3}^{2}} \,\, \sin (\theta)
\eeq
where $\theta  \sim \unif (0,2 \pi)$ and then $v_{2}$ by setting 
\beq
\elabel{v2sine2}
v_{2}^{\prime} = \sqrt{E-v_{3}^{2}-(v_{1}^{\prime})^{2}}.
\eeq
Four different values for $(v_{1},v_{2},v_{3})$ are visualized as points on the sphere in the first octant in Figure \ref{fig:spherepoints}, as well as an angle $\theta$ drawn uniformly over $(0, 2 \pi)$. All four points, {\bf{and indeed all points on the entire surface}}, will map onto the depicted arc.

\begin{figure}
\caption{Arc from Figure \ref{fig:spherepoints} mapped to shorter arc.}
\label{fig:spherearc}
\begin{subfigure}{0.5\textwidth}
\includegraphics[width=2.8in,height=2.6in]{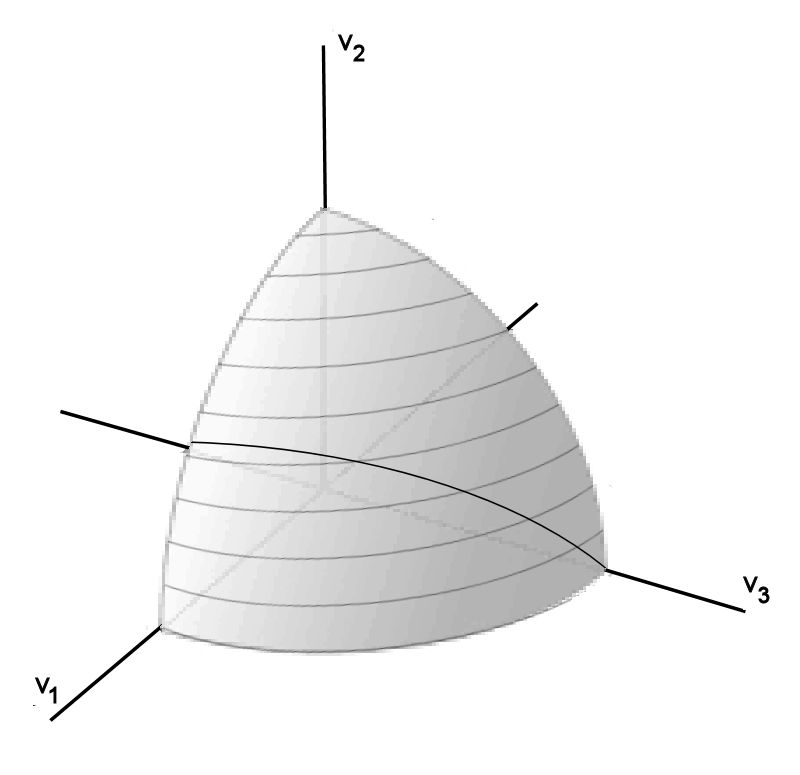}
\caption{Previous arc with rotated axes}
\end{subfigure}
\hspace{0.4in}
\begin{subfigure}{0.5\textwidth}
\includegraphics[width=2.8in,height=2.6in]{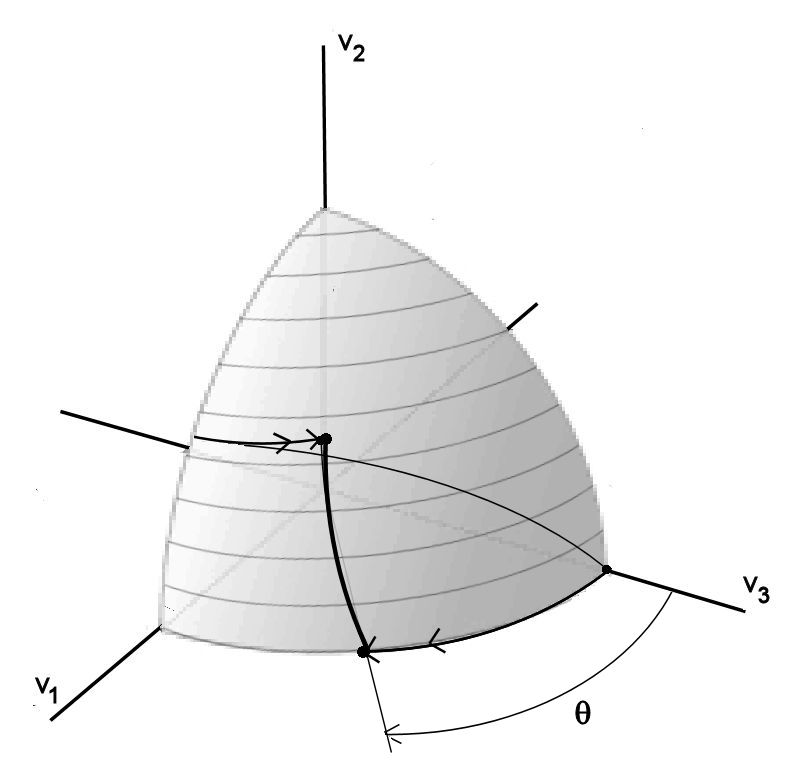}
\caption{New angle and arc mapped to smaller arc}
\label{fig:shb}
\end{subfigure}
\end{figure}

Suppose now that the next particles randomly selected for updating are indexed by $(i,j)=(3,1)$. Given a new angle $\theta$ drawn uniformly over $(0,2 \pi)$, the points on the arc from Figure \ref{fig:spherepoints} will map to the smaller arc segment depicted in Figure \ref{fig:spherearc}.

It is easy to see that if the same indices (in the same or different order) are selected twice in a row, the arc will maintain its length and move ``laterally''. However, if at least one index is different from those in the previous step, it is easy to verify that the arc segment will, necessarily, shrink in length. In other words, we will be mapping all points on the sphere closer and closer together. Thus, the algorithm is closing in on a single point on the surface. To formalize the $\varepsilon$-perfect sampling algorithm for general $N$, we will follow the $N$ corner points $\vec{c}_{1},\vec{c}_{2}, \ldots, \vec{c}_{N}$, where $\vec{c}_{i} = \sqrt{E} \, \vec{e}_{i}$ and $\vec{e}_{i}$ is the $i$th standard basis vector for $\field{R}^{N}$.

\vspace{0.2in}
{\bf{$\varepsilon$-Perfect Algorithm}}

Set an error tolerance $\varepsilon>0$ and set $n=1$.
\begin{enumerate}
\item Simulate and store $\theta_{-n} \sim \unif (0, \pi/2)$.

Randomly select a pair of distinct indices from $\{1,2, \ldots, N\}$ and store as a $2 \times 1$ vector $\vec{P}_{-n}$ with entries denoted as $\vec{P}_{-n}(1)$ and $\vec{P}_{-n}(2)$. (Order is important. For example $\vec{P}_{-n} = (3,5)$ is different from $\vec{P}_{-n}=(5,3)$.)

Set $\vec{c}_{i} = \sqrt{E} \, \vec{e}_{i}$ for $i=1,2,\ldots, N$.

Let $t=n$.

\item Let $\vec{c}_{i}(j)$ denote the $j$th element of $\vec{c}_{i}$.

For $i=1,2,\ldots, N$, let $e=[c_{i}(P_{-t}(1))]^{2} + [c_{i}(P_{-t}(2))]^{2}$ and set
$$
\vec{c}_{i}(P_{-t}(1)) = \sqrt{e} \,\, \sin \theta_{-t}
$$
and
$$
\vec{c}_{i}(P_{-t}(2)) = \sqrt{e-[\vec{c}_{i}(P_{-t}(1))]^{2}}
$$

If $t=1$, go to Step 3. Otherwise, let $t = t-1$ and return to the beginning of Step 2.

\item Compute the Euclidean distance between all pairs of corner points. Let $D$ be the maximum of these distances. If $D<\varepsilon$, compute the average corner point
$$
\vec{c}:=\frac{1}{N} \sum_{i=1}^{n} \vec{c_{i}},
$$
multiply each coordinate of $\vec{c}$ by $-1$ with probability $1/2$ independent of all other coordinates, output the new $\vec{c}$,
and exit the algorithm. Otherwise, set $n=n+1$ and return to Step $1$.

\end{enumerate}

In practice, it is more efficient, in the search for a backward coupling time,  to step back further than one step at a time. It is generally advised to let $t=2^{n}$ in, what is in this case, Step 1 of the $\varepsilon$-perfect algorithm. For simplicity, we use the Euclidean distance between corner points as a surrogate for arc length distance on the surface of the sphere since convergence to zero of the former implies convergence to zero of the latter.

\begin{figure}
\caption{Histogram of $100,000$ draws of $v_{1}$ Using the $\varepsilon$-perfect algorithm with $N=50$ and $\varepsilon= 10^{-6}$.}
\label{fig:spherev1}
\begin{center}
\includegraphics[width=4in,height=3.3in]{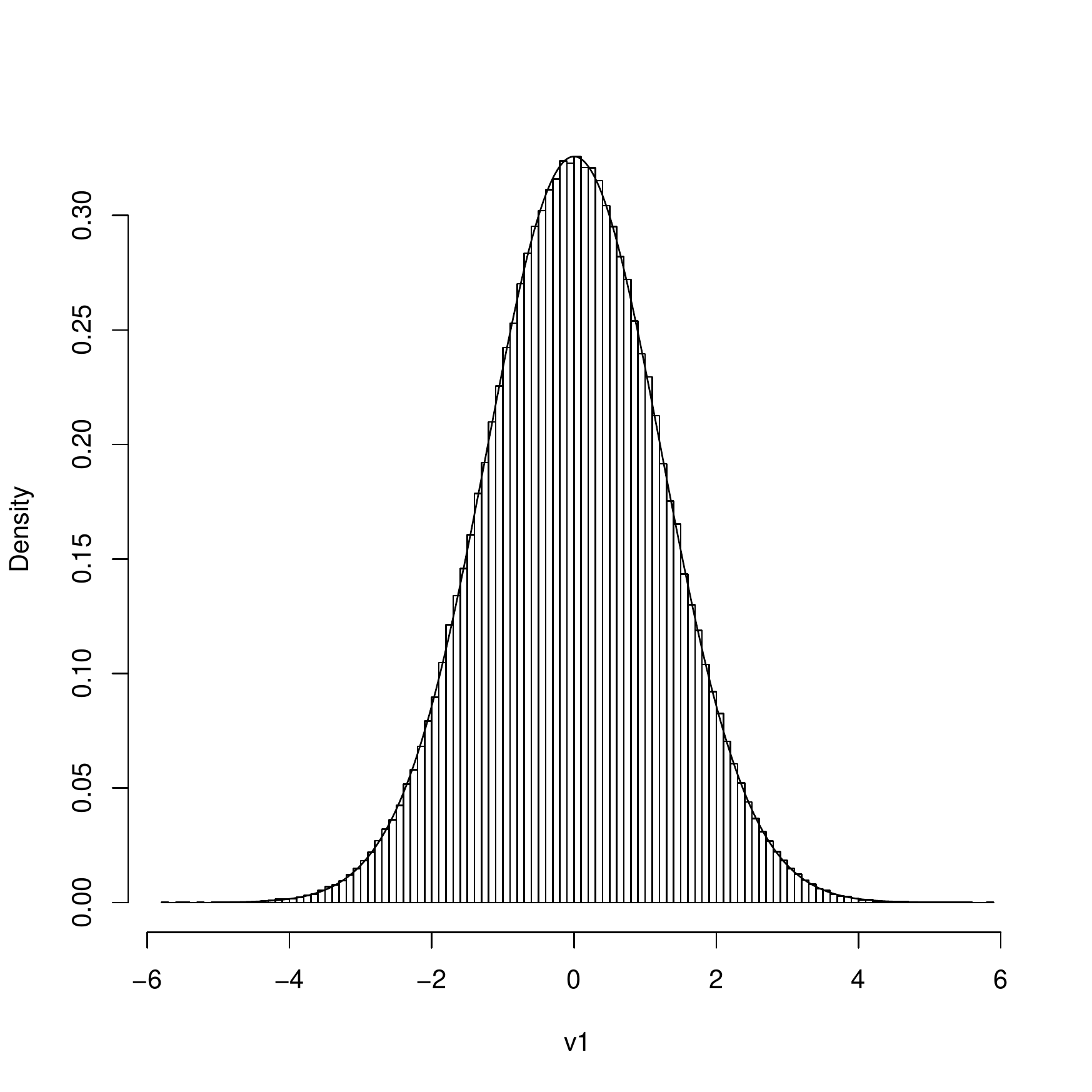}
\end{center}
\end{figure}

In Figure \ref{fig:spherev1} we show the results for $100,000$ draws of $v_{1}$ using the $\varepsilon$-perfect algorithm with  $N=50$ and $\varepsilon = 10^{-6}$ along with the curve given by (\ref{e:onedimkacsolinf}). For this simulation run, the mean backward coupling time was $948.2$ with a minimum of $422$ and a maximum of $1811$.

\section{Conclusions}
\slabel{conclusions}

In this paper we leveraged fundamental properties of the Poisson process to simulate values drawn from the probability density function that is the solution to the Kac equation in an efficient way that avoids the  tuning parameter $\Delta t$ that is necessary for the Nanbu and Nanbu-Babovsky algorithms. Since our algorithm takes advantage of the fact that the particle $1$ velocity is unchanged after a point in $(0,t)$, it expects to process a slightly lower number of collisions than Bird's DSMC algorithm. Furthermore, it appears to be performing better at capturing the tails of the distribution than the DSMC algorithm. This is likely due to the fact that we have a variable Poisson number of collisions in any time  interval as opposed to the constant expected value.

We also introduced, to our knowledge, the first perfect sampling type algorithm to be applied to the Kac model as a proof of concept that has the potential to be expanded to more interesting applications that are currently simulated using variants of the DSMC algorithm. A truly perfect ($\varepsilon=0$) algorithm using a minorization approach (as in \cite{cortwe01}) is possible but the increase in accuracy is not worth added complication, especially in light of the fact that we can set $\varepsilon$ in line with the maximum attainable computer dependent precision.


\bibliographystyle{plain}

\bibliography{kac_poisson_arxiv}

\end{document}